\documentclass[english,preprint]{svjour3}
\usepackage{mathptmx}
\usepackage[T1]{fontenc}
\usepackage{geometry}
\usepackage{babel}
\usepackage{units}
\usepackage{url}
\usepackage{amssymb}
\usepackage{graphicx}
\usepackage[authoryear]{natbib}
\usepackage[unicode=true]
 {hyperref}

\makeatletter

\providecommand{\tabularnewline}{\\}

\makeatother

\begin{document}

\title{Empirical evidences for a planetary modulation of total solar irradiance
and the TSI signature of the 1.09-year Earth-Jupiter conjunction cycle}

\author{Nicola Scafetta$^{1,2}$ and Richard C. Willson$^{1}$}

\institute{$^{1}$Active Cavity Radiometer Irradiance Monitor (ACRIM), Coronado,
CA USA 92118}

\institute{$^{2}$Duke University, Durham, NC 27708}
\maketitle
\begin{abstract}
The time series of total solar irradiance (TSI) satellite observations
since 1978 provided by ACRIM and PMOD TSI composites are studied.
We find empirical evidence for planetary-induced forcing and modulation
of solar activity. Power spectra and direct data pattern analysis
reveal a clear signature of the 1.09-year Earth-Jupiter conjunction
cycle, in particular during solar cycle 23 maximum. This appears to
suggest that the Jupiter side of the Sun is slightly brighter during
solar maxima. The effect is observed when the Earth crosses the Sun-Jupiter
conjunction line every 1.09 years. Multiple spectral peaks are observed
in the TSI records that are coherent with known planetary harmonics
such as the spring, orbital and synodic periods among Mercury, Venus,
Earth and Jupiter: the Mercury-Venus spring-tidal cycle (0.20 year);
the Mercury orbital cycle (0.24 year); the Venus-Jupiter spring-tidal
cycle (0.32 year); the Venus-Mercury synodic cycle (0.40 year); the
Venus-Jupiter synodic cycle (0.65 year); and the Venus-Earth spring
tidal cycle (0.80 year). Strong evidence is also found for a 0.5-year
TSI cycle that could be driven by the Earth's crossing the solar equatorial
plane twice a year and may indicate a latitudinal solar-luminosity
asymmetry. Because both spring and synodic planetary cycles appear
to be present and the amplitudes of their TSI signatures appear enhanced
during sunspot cycle maxima, we conjecture that on annual and sub-annual
scales both gravitational and electro-magnetic planet-sun interactions
and internal non-linear feedbacks may be modulating solar activity.
Gravitational tidal forces should mostly stress spring cycles while
electro-magnetic forces could be linked to the solar wobbling dynamics,
and would mostly stress the synodic cycles. The observed statistical
coherence between the TSI records and the planetary harmonics is confirmed
by three alternative tests. 
\end{abstract}

\section{Introduction}

Numerous observations -- e.g. sunspot, total solar irradiance (TSI)
satellite \citep{Willson2003,Frohlich} and magnetic flux records
\citep{Ball} -- have demonstrated that solar activity is characterized
by a variable $\sim$11-year Schwabe cycle, by complex dynamics on
monthly-to-annual time scales and by possible multi-decadal trending.
In addition to a varying 11-year solar cycle, spectral analyses of
solar records have also identified a number of distinct periodicities
at the multidecadal, secular and millennial scales \citep{Abreu,Bond,Frick,Ogurtsov,Scafetta2012c,Scafetta2013,Tan2011}.
The statistically estimated periodicities depend on the length and
nature of the analyzed records. However, typical major spectral peak
values are found at 43-45 years, 55-61 years, 81-87 years (Gleissberg),
98-130 years, 150-180 year, $\sim$207 years (de Vries), $\sim$500
years, $\sim$980 years (Eddy), and others. See for example \citet{Scafetta2012c}
and \citet{Scafetta2013} where high resolution spectral analyses
of a 438-year long auroral record and of a 9,400 solar proxy model
\citep{Steinhilber} are proposed. 

The conventional view of solar science has been that solar magnetic
and radiant variability are driven by internal solar dynamics alone,
characterized by hydromagnetic solar dynamo models \citep{Tobias,Jiang}.
However, current solar dynamo models assume that the Sun is an isolated
system and have been unable to explain and/or forecast solar variability
including the emergence of the Schwabe 11-year cycle and of its multidecadal,
secular and millennia modulations. In fact, although solar dynamo
theories do predict the emergence of generic solar oscillations, the
reason why, among all possible theoretical frequencies, the Sun oscillates
at the specific observed frequencies and phases has eluded researchers
so far. For example, periods of low solar activity such as the Maunder
and Dalton minima, and the intra-annual oscillations larger than the
monthly solar differential rotation cycles are essentially not predicted
by the solar dynamo models, which require an arbitrary choice of specific
free parameters even for reproducing an 11-year major oscillation
\citep{Jiang}.

An alternative theory for solar variability was first put forth in
the 19th century by numerous auroral and solar specialists such as
\citet{Wolf}, and investigated later by \citet{Brown}, \citet{Luby}
and others. It proposes that solar magnetic and radiant variability
is partially regulated by planetary gravitational and magnetic forcings.
This theory was originally suggested by the related facts that: (1)
sunspot activity was found to be characterized by an $\sim$11-year
cycle; and (2) the Schwabe frequency band falls between the two primary
tidal cycles associated with the Jupiter-Saturn spring tide (9.93-year)
and with Jupiter's orbital period (11.86-year). \citet{Bendandi}
observed that the tides of Venus, Earth and Jupiter generate a recurrent
oscillation at the period of 11.07 years and \citet{Wood} observed
that the solar-jerk function presents a period of 11.08 years. Indeed,
the average 11-year solar cycle length is 11.05-11.10 years \citep{Scafetta2012c}.
These results have been more recently confirmed and expanded \citep[e.g.:][]{Hung,Scafetta2012c,Scafetta2012d}.
Thus, internal solar dynamo mechanisms may be synchronized to the
frequencies of the external planetary harmonic forcings. 

The planetary theory of solar variation has not found support within
the mainstream solar physics community during the last decades because:
(1) it was found problematic to qualitatively relate secular solar
variability to planetary cycles \citep{Smythe}; (2) planetary forcings
appear weak over the sun-planet distances \citep{Callebaut}; (3)
although planetary induced solar luminosity enhancement observations
have been reported for numerous extrasolar planets \citep{Scharf,Gurdemir,Wright}
interpretation of the data referring to distant solar systems remains
controversial \citep{Poppenhaeger}. While some cyclical variability
has been observed in numerous sun-like stars \citep{Baliunas} unambiguous
planetary influence cannot be deduced since the planetary configuration
of those distant solar systems is mostly unknown, stellar data are
available only for a few decades at most and their quality may be
still too poor.

The planetary configuration of our solar system is known very well
and there are solar data and magnetic activity proxy models spanning
decades to millennia that are of relatively high quality. A number
of empirical studies have found preliminary evidence for planetary
influences on solar behavior \citep{Bigg,Charvatova,Cionco,Fairbridge,Hung,Jose,Juckett2000,Juckett2003,Landscheidt,Leal,Sharp,Wilson}.
Spectral analysis has recently revealed that the sunspot Schwabe cycle
may be made and modulated by at least three frequencies \citep{Scafetta2012c}:
the two Jupiter and Saturn tidal cycles (9.93 and 11.86 year) and
the strong central 10.87-year dynamo cycle, which may also emerge
from other planetary recurrent cycles varying between 10.40 and 11.10
years due to Venus, Earth and Jupiter. A harmonic model based on these
three major frequencies suffices to hindcast the major solar secular
solar oscillation (e.g. the Oort, Wolf, Sp\"orer, Maunder and Dalton
minima) and the millennial solar oscillation as interference patterns
produced by the three cycles throughout the Holocene. Thus, if planetary
harmonics are synchronizing and modulating the solar dynamo, specific
major oscillations (quasi 11-year, 20-year, 42-45-year, 61-year, 86-year,
115-year, 130-year, 170-180-year, 207-year, 500-year and 1000-year
cycles) would characterize solar activity. Indeed, these oscillations
are commonly observed in both solar and the Earth's climate records
\citep{Bond,Ogurtsov,Qian,Chylek,Steinhilber,Scafetta2012c,Scafetta2012d}.
Other recent studies have noted planetary harmonics in the sunspot
record, auroral records, microwave emission and multi-millenia cosmo-nucleotide
proxy solar models \citep{Abreu,Scafetta2012a,Scafetta2013,Tan}.

The physical problem is not solved yet and is highly controversial.
However, preliminary gravitational physical mechanisms have been recently
proposed \citep{Wolff,Scafetta2012d,Abreu}. Internal solar feedback
processes could also lead to increased magneto-acoustic heating, dynamo
action and direct magnetic interactions between stellar and planetary
magnetic fields \citep[e.g:][]{Cuntz,Scharf}. Alternative physical
mechanisms may also simultaneously act and/or be complementary to
each other. For example, gravitational tides and electromagnetic mechanisms
may generate alternative sets of harmonics with the former stressing
the spring harmonics and the latter the synodic harmonics among the
planets.

\citet{Wolff} argued that planetary gravitational forcing could cause
a mass flow inside the Sun that could carry fresh hydrogen fuel to
deeper levels including the solar core consequently increasing the
solar nuclear fusion rate. \citet{Abreu} proposed that planetary
tides could exert a varying torque on a non-spherical tachocline,
perturb the operation of the solar dynamo and modulate the long-term
solar magnetic activity. \citet{Abreu} found that the torque signal
presents numerous harmonics on the secular and multisecular scales
in common with long solar proxy models. In general, a frequency matching
between solar and planetary harmonics may be achieved adopting alternative
physical functions of the orbits of the planets and, therefore, it
would suggest the existence of a physical link between planetary orbits
and solar activity, without necessarily specifying the exact physical
mechanisms involved in the process.

\citet{Scafetta2012d} observed that planetary gravitational tidal
forcing alone appears too weak to directly force the solar tachocline
or the solar convection zone. \citet{Callebaut} dismissed a possible
direct planetary forcing of the solar convection zone by arguing that
the internal convection velocities and accelerations are far greater
than those derivable from planetary tidal influences. The claim needs
to be partially reconsidered because \citet{Shravan} have recently
found that the solar convection velocities appear to be 20-100 times
weaker than the previous theoretical estimations. \citet{Scafettaal2013}
have shown other shortcomings of the Callebaut et al. analysis. In
any case, it appears that planetary forcings may modulate solar activity
only if internal solar mechanisms greatly amplify their effects. 

\citet{Scafetta2012d} proposed that the gravitational energy released
by the planetary tides to the sun may trigger slight nuclear fusion
rate variations by enhancing solar plasma mixing. In fact, solar plasma
is made of protons and electrons that can freely move and interact
through electromagnetic forces. Under gravitational perturbations
electrons and protons may drift in opposite directions perpendicular
to the gravitational forces generating micro currents in the plasma.
For example, in a magnetic field $\vec{B}$ a plasma moved by a local
gravitational field $\vec{g}$ has its protons and electrons drifting
with a velocity given by $\vec{v_{g}}=m\vec{g}\times\vec{B}/qB^{2}$,
where the dependence on the charge $q$ of the particle implies that
the drift direction is opposite for protons and electrons, resulting
in an electric current. This relative movement of electrons and protons
may cause a local reduction of the electric repulsion among protons,
which can enhance and modulate the nuclear fusion rate in proportion
to the strength of the gravitational tides acting in the core. A nuclear
fusion enhancement can yield a significant amplification of the gravitational
tidal energetic signal released to the sun. 

\citet{Scafetta2012d} observed that a gravitational triggering of
the nuclear fusion rate could be possible because main-sequence stars
are characterized by a delicate balance between nuclear fusion activity
and gravitational forces. Essentially, in stars gravity triggers nuclear
fusion activity and controls it, and every gravitational variation
inside the star could trigger a variation of the luminosity production.
Scafetta proposed a physical theory to quantify the phenomenon based
on the stellar mass-luminosity relation: $L_{1}/L_{2}\sim(M_{1}/M_{2})^{4}$
\citep{Duric}. The stellar mass-luminosity relation may also be interpreted
as a relation between a star's luminosity and the gravitational work
released to the star by solar gravitational forces, which are related
to the solar mass. For example, if the mass of the planets of the
solar system were added to the Sun, the solar luminosity should increase
by about 0.5\%, and TSI would increase by about $7.4$ $W/m^{2}$. 

The nuclear fusion amplification mechanism proposed by \citet{Scafetta2012d}
causes an amplification of the tidal signal through nuclear fusion
enhancement of the order of $A\approx4\cdot10^{6}$, which can produce
a small tidal-induced luminosity variation comparable with that of
the observed TSI fluctuations, that is up to 0.05-1 $W/m^{2}$ \citep{Willson1986,Willson1991,Willson2003}.
If Scafetta's proposal is correct, a tidal induced luminosity signal
may be sufficiently strong to synchronize the operation of the solar
dynamo. For example, it would produce a sufficiently strong planetary
tidal induced solar bulge on a $\sim6^{o}$ inclined plane (the solar
system disk), relative to the solar equatorial bulge, causing a varying
torquing force on the solar convection zone \citep[see also][]{Luby}.
\citet{Scafetta2012d}'s theoretical model predicts the two tidal
cycles ($\sim$9.93 years and $\sim$11.86 years) associated with
Jupiter and Saturn observed in the sunspot record as well as significant
complex intra-annual fluctuations. The complex intra-annual oscillation
would be dominated by the $\sim$0.32-year Jupiter/Venus spring tide
oscillation and by numerous other specific planetary frequencies.
\citet{Scafetta2012d} also noted that the pattern of these fluctuations
approximately repeats every 11.08 years. This period corresponds to
the average Schwabe solar cycle length \citep{Scafetta2012c}. 

In response to typical reservations expressed about gravitational
triggering of the nuclear fusion rate, namely that photons diffuse
very slowly from the core to the tachocline (e.g., the Kelvin\textendash{}Helmholtz
timescale for the Sun may be as long as $10^{4}$-$10^{5}$ years)
and any core luminosity fluctuation would be damped before reaching
the tachocline, \citet{Wolff} and \citet{Scafetta2012c,Scafetta2012d}
postulated the existence of fast acoustic-like or g-wave transport
mechanisms. Wave mechanisms could transport an energy variation signal
from the core to the tachocline within a time scale of a few weeks.
Essentially, as the core luminosity production varies, the core would
expand and contract in function of the strength of the planetary tides
and this perturbation should be felt by the entire sun quite fast.
Once at the tachocline, the small harmonic signal emerging from the
radiative zone would force the convection zone, synchronize the operation
of the solar dynamo and, finally, modulate the solar luminosity outputs.
It may be conjectured that a gravitational planetary modulation of
the solar luminosity production could be demonstrated by a synchronized
variation of the solar neutrino production. However, a test of this
hypothesis is not currently possible because the statistical uncertainty
of neutrino measurements is too large (up to 20\%-50\% of the observed
value) \citep{Sturrock}, while the magnitude of \citet{Scafetta2012d}'s
hypothesized planetary-induced luminosity variation is about 0.05\%-0.005\%
of the solar luminosity. 

A final important observation is that even if planetary forcings are
modulating solar activity the magnitudes of their effects on solar
observables cannot be estimated with precision because the exact physical
forcing mechanisms (gravitational and electromagnetic) and internal
feedbacks -- e.g. resonances and synchronization mechanisms \citep{Pikovsky}
that may produce internal dynamical amplifications but also prevent
the formation of too regular harmonics characterized by sharp spectral
peaks -- remain controversial and largely unknown. However, if planetary
forcings are modulating solar activity, their effects should generate
a complex set of harmonics associated with the periods of the planetary
orbits and their mutual spring and synodic cycles. Thus, specific
frequencies bands of the records describing solar activity are expected
to be activated by planetary forcings. As an analogy, in the case
of the ocean tides on the Earth only theoretical frequencies can be
carefully calculated from astronomical considerations while the actual
amplitudes and phases of the single harmonics can only be directly
measured from tide gauge records because the exact physical mechanisms
generating them are not known with precision \citep{Kelvin,Doodson,Wang}.
Therefore, a planetary modulation of solar activity can be demonstrated
by studying the spectral harmonics of solar activity records and by
checking whether a common set of frequencies between the solar records
and the theoretical planetary frequencies exists. 

Herein we conjecture that planetary harmonic signatures should be
observed in total solar irradiance (TSI) satellite observations on
the monthly-to-annual scales too as shown in a recent conference presentation
by \citet{Scafetta2010b}. Thus, herein we study whether the ACRIM
and PMOD TSI satellite composites (both continuous in TSI since 1978)
present signatures of planetary harmonics.

\section{TSI data and harmonic orbital expectations}

Composites of TSI satellite observations are continuous since late
1978. The two most cited composites are the ACRIM \citep{Willson2003}
and PMOD \citep{Frohlich}. Before mid-1992 ACRIM and the PMOD disagree
about the quality and use of the extant TSI data, and the composites
disagree in important ways \citep{Frohlich,Scafetta2009,Scafetta2011}.
\citet{Frohlich} alters NIMBUS7/ERB and ACRIM1 published records
\citep{Willson1981,Willson1991,Willson1997} for reasons that are
disputed by the original science teams \citep{Willson2003,Scafetta2009}.
Because of the data controversy we will confine our spectral analysis
to the less disputed ACRIM data for mid-1992 to 2012. In this range
the ACRIM composite is comprised of the results from the ACRIM2 (1992-2000)
and ACRIM3 (2000-2012) experiments. During this same period the PMOD
composite is comprised of the ACRIM2 (1992-1996 and part of the period
1998-1999) and VIRGO results (most of 1996-2012). Since variable data
patterns are amplified during solar active periods, we compare ACRIM
and PMOD during the maximum of solar cycle 23 (1997.75-2004.25).

Because we are observing the Sun from satellites orbiting the Earth
the major cycle found in TSI satellite raw data is related to the
1-year Earth's orbital cycle. TSI records are carefully rescaled to
1 AU using accurate ephemerides, corrected for sensor degradation
and temperature sensitivities. A rigorous 1-year cycle, if found,
could be due to uncorrected sensor artifacts and/or to a latitudinal
solar luminosity asymmetry. If an annual cycle is found it should
be carefully analyzed and accounted for, if necessary. 

The strongest theoretical planetary induced tides on the Sun are produced
by Jupiter, followed by Venus, Earth and Mercury, in order of magnitude
and their values are calculated in \citet{Scafetta2012d}. If electro-magnetic
interactions throughout the Parker heliospheric current spiral exist,
Jupiter would still be the planet with the greatest effect because
it has the largest magnetosphere. Earth and Venus would influence
the high frequency component of the spectrum, while the other Jovian
planets (Saturn, Uranus and Neptune) could more likely influence the
low multidecadal and multisecular variability \citep{Abreu,Scafetta2012c,Scafetta2013,Sharp}. 

Herein we are looking for high frequency harmonics with period up
to about $1.2$ years. The gravitational tidal cycles stress the spring
tides among the planets. Other mechanisms (e.g electromagnetic couplings)
would be superimposed upon the tidal cycles and would stress the planetary
synodic/conjunction cycles. Electromagnetic mechanisms may be related
to the speed and jerk functions of the Sun relative the barycenter
of the solar system \citep{Scafetta2010,Scafetta2012d}. Specific
torque components acting on the solar masses could be related to the
z-axis of the tidal forcings. 

If TSI is modulated by planetary forces the records should be characterized
by numerous planetary harmonics. The most relevant periods involve
the four major tidal planets: Mercury, Venus, Earth and Jupiter (see
Table 1). Table 2 lists additional spring and synodic periods using
the four minor tidal planets: Mars, Saturn, Uranus and Neptune. Given
the periods of two planets $P_{A}$ and $P_{B}$, their synodic period
$P_{AB}$, which is twice the spring period, is defined as 

\begin{equation}
P_{AB}=\frac{1}{|1/P_{A}-1/P_{B}|}.\label{eq:1}
\end{equation}

A major cycle is anticipated for the strong Venus-Jupiter spring tidal
cycle with a periodicity of $\sim$0.32-year. A 0.5-year periodicity
TSI harmonic could be produced by the Earth\textquoteright{}s crossing
the solar equator twice a year, providing a changing view of the northern
and southern solar hemispheres. If the Jupiter side of the Sun is
the brightest one, the Earth-Jupiter synodic cycle should cause a
TSI cycle with a $\sim$1.092-year periodicity.

Table 3 lists addition relevant harmonics. These include the solar
differential equatorial and polar rotation cycles relative to the
fixed stars and to the four major tidal planets. We use an adaptation
of Eq. \ref{eq:1} where one planetary harmonic is substituted with
the solar rotation periods. Table 3 also lists the spring and synodic
periods of Mercury, Venus and Earth relative to the synodic periods
of Jupiter and Saturn, and Jupiter and Earth using the three-synodic
equation: 

\begin{equation}
P_{A(BC)}=\frac{1}{|1/P_{A}-|1/P_{B}-1/P_{C}||}.\label{eq:1-1}
\end{equation}
Other harmonics related to specific planetary configurations are possible,
but are ignored herein.

\section{Spectral Coherence between TSI records and planetary harmonics}

Spectral analysis results are shown in Figure 1. Figure 1A depicts
the traditional Lomb \citep[unevenly sampled data:][]{Press} periodogram
of the 1992.5-2012.9 ACRIM TSI record. The statistical confidence
of the frequency peaks is tested against red-noise background at 95\%
and 99\% levels. Moreover, the standard deviation of the data within
the analyzed frequency band (for periods less than 1.2 year), $\sigma_{d}=0.34$
$W/m^{2}$, is 26 times larger than the average instrumental statistical
error ($\sigma_{i}=0.013$ $W/m^{2}$). So, the observed TSI frequency
patterns cannot be due to random uncertainties in the measurements.

Figure 1B uses the maximum entropy methodology \citep[MEM: ][]{Press}
in the traditional manner \citep{Courtillot,Scafetta2012b} to reanalyze
the ACRIM record and compare it with the PMOD record during solar
cycle 23 activity maximum (1997.75-2004.25) when PMOD contains mostly
non-ACRIM data. The results depicted in Figures 1A and 1B were derived
using different methodologies, data and time intervals. They validate
each other because a common set of statistically significant spectral
peaks is found at about 0.32 year, 0.5 year and 1.09 year.

Figures 1A and 1B also depict the major frequency bands (yellow boxes)
associated with theoretical planetary harmonics deduced from the spring,
orbital and synodic periods among the major tidal planets (Mercury,
Venus, Earth and Jupiter): see Tables 1, 2 and 3. We assume an average
$\pm5$\% variability in the periods which is caused by three factors.
The first is the elliptical shape of planetary orbits that make spring
and synodic periods vary while the theoretical orbital periods do
not vary significantly (Mercury-Venus spring period varies as $0.198\pm0.021$
year, Earth-Jupiter synodic period varies as $1.092\pm0.009$ year,
Venus-Earth spring period varies as $0.799\pm0.008$ year: see Table
1). Second, there may be an internal nonlinear solar response that
makes the TSI periodicities fluctuate around the theoretical forcing
frequencies. Third, there are statistical errors in the data and in
the power spectrum evaluations. 

The yellow bar histogram shown in Figure 1A and 1B provides a best
guess representation of the planetary spectrum constructed using bars
centered on the planetary frequencies listed in the tables with a
$\pm5\%$ width. The relative heights of the bars take into account:
(1) a possible relative strength calculated using the weight $d^{-3}$
where $d$ is the distance a planet from the Sun (the greatest effect
is given by Mercury followed by Venus, Earth and Jupiter); (2) the
mass $M$ of the planet (the greatest effect is given by Jupiter,
followed by Earth, Venus and Mercury); (3) the strength of the planetary
magnetosphere (the greatest effect is given by Jupiter, followed by
Earth, Venus and Mercury); (4) spring and synodic oscillations should
be stronger than orbital periods; and (5) symmetries associated with
the Earth's orbit due to the fact that TSI is defined as the solar
irradiance radiated toward the Earth at 1 astronomical unit from the
Sun. Finally, analogous to \citet{Kelvin}'s ocean tidal method that
determines the tidal amplitudes from tide gauge records, the heights
of the bars are chosen to reproduce the correspondent heights observed
in the peaks of the TSI power spectrum functions. Our choice of the
heights of the yellow bars should be considered best guesses due to
the fact that, as explained above, it is not possible at the moment
to deduce the TSI harmonic amplitudes from physical principles regulating
planet-sun interactions. 

Figure 1A and 1B show that major TSI spectral peaks are found within
our spectral bands at 0.30-0.33 years, 0.5 years and 1.09 years that
characterize major expected planetary harmonics. Other expected frequency
clusters within our spectral bands at 0.2 year, 0.25 year, 0.4 year,
0.6-0.7 year and 0.9-1 year are observed. This result is particularly
evident in Figure 1A where the longer TSI record from mid-1992 to
2012 is used. Thus, a simple visual analysis between the TSI spectra
and the expected planetary-induced harmonics depicted in Figure 1
already suggests that planetary forcings modulate TSI. It should be
noticed that a rigorous 1-year cycle is not seen in the analysis reported
in the figure.

\begin{figure}
\begin{centering}
\includegraphics[width=1\textwidth,height=0.7\textheight,keepaspectratio]{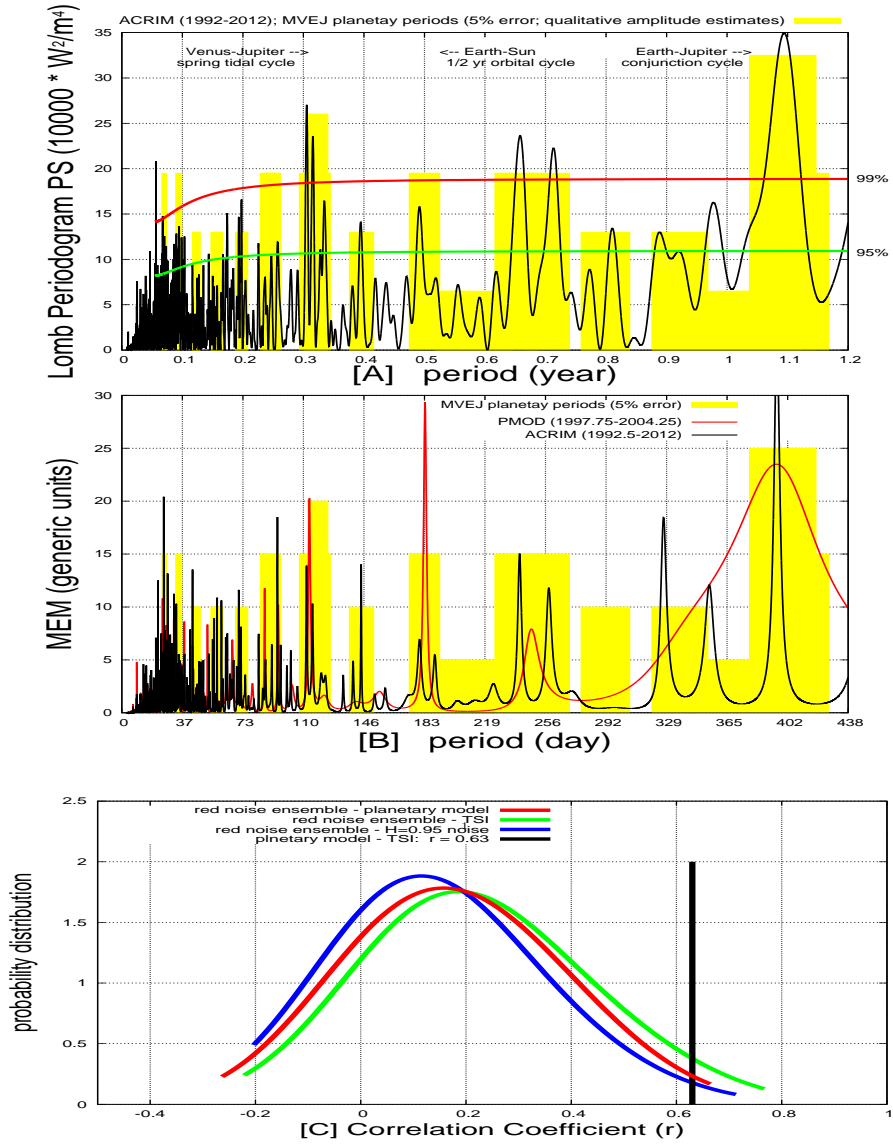}
\par\end{centering}

\caption{{[}A{]} Periodogram of ACRIM results in $W^{2}/m^{4}$ units for data
from 1992.5-2012, against red noise background at 95\% and 99\% confidence
level \citep{Ghil}. {[}B{]} Power spectrum based on the Maximum Entropy
Method of ACRIM (1992.5-2012.9) and PMOD during the maximum of solar
cycle 23 (1997.75-2004.25). The yellow bars are a representation of
the planetary spectrum due to the harmonics generated by the planets
reported in Tables 1-3. {[}C{]} Monte Carlo tests to determine the
statistical confidence of the correlation coefficient between the
yellow area representing a spectral model of the planetary oscillations
and the observed TSI periodogram of figure A ($r\thickapprox0.63$;
black). }
\end{figure}

\subsection{Test \#1: Monte Carlo red-noise simulation}

Here we present two different tests to evaluate the coherence between
the TSI spectrum and the planetary harmonics. The first test is depicted
in Figure 1C and it is based on a Monte Carlo simulation for determining
the statistical significance of the spectral correlation between the
TSI periodogram and the yellow schematic planetary spectral model
in Figure 1A, which represents our physical hypothesis. The found
correlation value is $r\thickapprox0.63$ and is indicated by a black
vertical bar in Figure 1C. The test is run against the null hypothesis
that the TSI record may be a red-noise signal. This test aims to determine
how easy would be to find a surrogate red-noise random signal characterized
by a periodogram correlated with the planetary spectrum model (indicated
by the yellow histogram) with a spectral coherence correlation coefficient
$r\gtrsim0.63$. 

We use red-noises because they are statistical autocorrelated signals
generated by typical dynamical fractal mechanisms. This is how self-regulating,
autocorrelated chaotic systems generally evolve and how solar activity
would be expected to evolve if a specific set of harmonic forces did
not regulate it. Our Monte Carlo test uses 15,000 red-noise sequences
as TSI surrogate records with different Hurst exponents, $H$, varying
between 0.85-1.05, corresponding to the range of the solar fractal
variability. The average Hurst exponent for TSI has been measured
to be $H=0.95$ \citep{Scafetta2003,Scafetta2004,Scafetta2005}. Thus,
we repeated the spectral analysis using these 15,000 different red-noise
records of length equal to the analyzed TSI data. The periodogram
of each TSI surrogate record is calculated and cross-correlated with
the yellow schematic planetary spectral curve representing our physical
hypothesis by repeating the same measure used for the original TSI
record. It is found that the 15,000 independent TSI synthetic records
produce periodograms that correlate with the yellow planetary function
with a spectral coherence correlation coefficient $r\leq0.66$ and
only $8$ red-noise simulations produced a $r$ value larger than
$0.63$. The probability distribution of these correlation coefficients,
that is the probability distribution of the null-hypothesis, is depicted
in Figure 1C, red curve. 

Thus, the probability to obtain the found correlation coefficient
between the real TSI periodogram and the yellow histogram schematically
modeling the planetary oscillations in Figure 1A (that is, $r\approx0.63$)
using generic red-noise surrogate TSI records is less than $0.05\%$.
This Monte Carlo statistical test supports the claim that the analyzed
TSI record presents dynamical spectral patterns that are better correlated
to planetary harmonics than to generic red-noises that can be produced
by autocorrelated signals typically generated by self evolving chaotic
systems. Figure 1C also adds probability distributions of correlation
coefficients between the red-noise ensemble directly with the TSI
spectrum (green) and with a generic red-noise with Hurst exponent
$H=0.95$ (blue), which is the average fractal exponent for TSI \citep{Scafetta2005}.
Also these two distributions suggest that the found correlation coefficient
between the planetary model and TSI, $r\thickapprox0.63$, is highly
significant statistically. 

\begin{figure}
\begin{centering}
\includegraphics[width=1\textwidth,height=0.8\textheight]{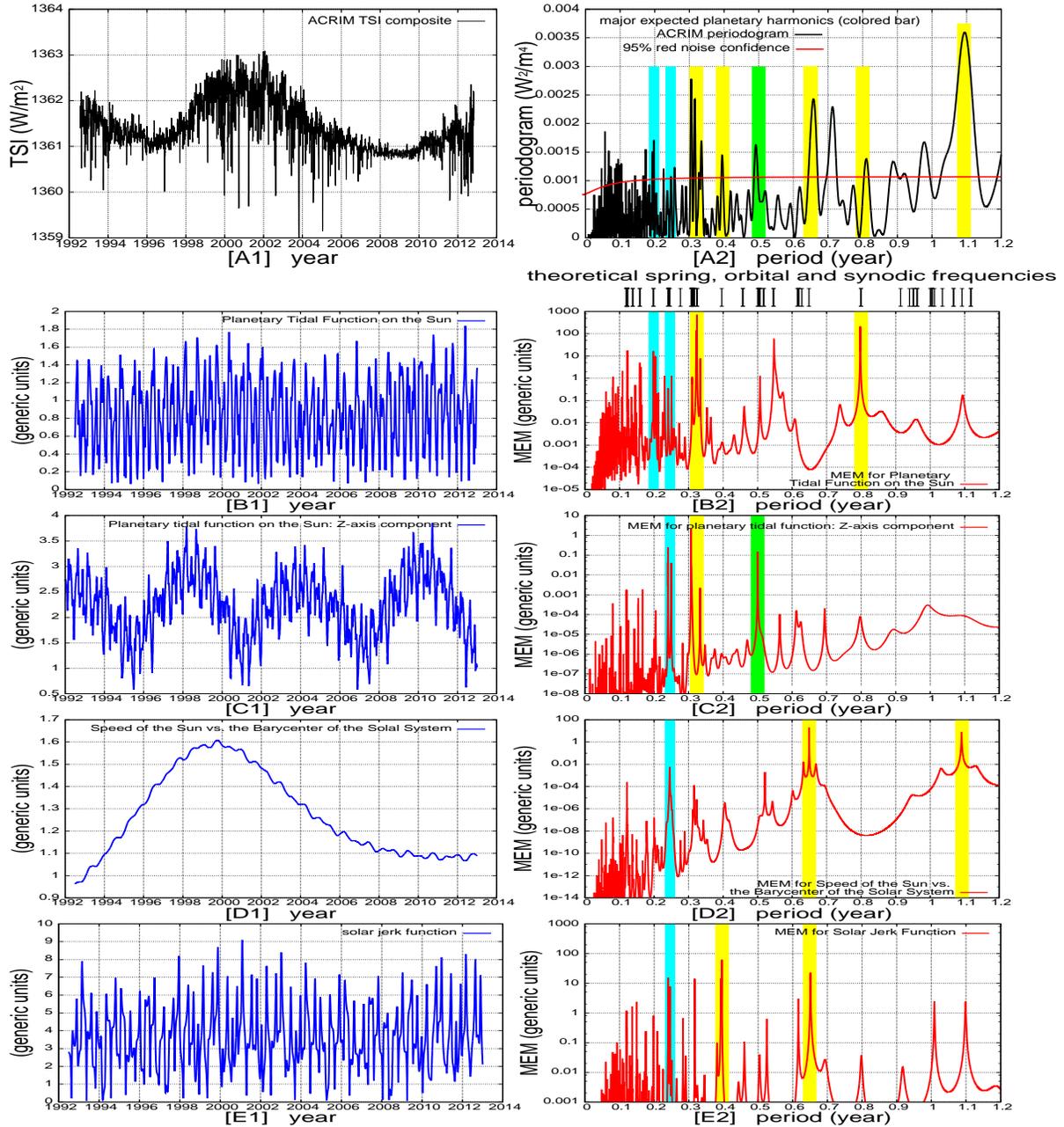}
\par\end{centering}

\caption{{[}A{]} Data (left) and periodogram (right) of the ACRIM TSI composite
from 1992 to 2013. Data (left, blue) and MEM power spectrum analysis
(right, red) of: {[}B{]} the tidal function produced by all eight
planets of the solar system (from Mercury to Neptune) on the Sun as
defined in \citet{Scafetta2012d}, see Eq. \ref{3B}; {[}C{]} the
the Z-axis component of the tidal function produced by all eight planets,
see Eq. 3; {[}D{]} the speed of the Sun relative to the barycenter
of the solar system as used in \citet{Scafetta2010} as calculated
by the JPL's HORIZONS system; {[}E{]} the jerk function (the derivative
of the acceleration vector) of the Sun as defined in \citet{Scafetta2012d},
see Eq. \ref{3E}. The colored bars indicate the largest significant
common peaks. The black small bars between panel A2 and B2 indicate
all theoretical spring, orbital and synodic periods expected from
the eight planets of the solar system within the analyzed period band
from 0 to 1.2 year: see Tables 1 and 2. The quasi 0.7 year cycle in
2A appears to be related to the spring period of Venus beating with
the synodic cycle of Earth and Jupiter (see Table 3).}
\end{figure}

\subsection{Test \#2: spectral coherence with planetary theoretical functions}

The second test, using the results depicted in Figure 2, aims to determine
at which planetary frequencies the strongest TSI response could be
expected from hypothesized basic physical principle. Figure 2A shows
the ACRIM TSI record and its periodogram as in Figure 1A. Following
the arguments of Section 2 regarding the physical origins of the involved
mechanisms, Figure 2 shows the MEM power spectra of the following
four records for the period 1992-2013. These are calculated using
JPL's HORIZONS ephemeris data downloaded from its web-interface (\href{http://ssd.jpl.nasa.gov/horizons.cgi}{http://ssd.jpl.nasa.gov/horizons.cgi})
for all solar system planets. 

Panel 2B shows the energy released by tides produced from the planets
on the Sun as defined in \citet{Scafetta2012d} that uses $m_{p}$for
the mass of the planet $p$ and $R_{PS}(t)$ for the sun-planet distance:

\begin{eqnarray} I_{P}(t) & = & \frac{3~G~R_{S}^{5}}{2~Q~\Delta t}\int_{0}^{1}K(\chi)~\chi^{4}\rho(\chi)~\texttt{d}\chi\cdot\label{eq912}\\  &  & \int_{\theta=0}^{\pi}\int_{\phi=0}^{2\pi}\left|\sum_{P=1}^{8}~m_{P}\frac{\cos^{2}(\alpha_{P,t})-\frac{1}{3}}{R_{SP}^{3}(t)}-~m_{P}\frac{\cos^{2}(\alpha_{P,t-\Delta t})-\frac{1}{3}}{R_{SP}^{3}(t-\Delta t)}\right|~\sin(\theta)~\texttt{d}\theta\texttt{d}\phi;\nonumber  \end{eqnarray}

Panel 2C shows the solar z-axix component function of the planetary
tidal accelerations relative to the solar equatorial reference plane,
defined as 

\begin{equation}
F_{z}(t)\propto\sum_{p=1}^{8}\frac{m_{p}|z_{p}(t)|}{R_{SP}^{4}(t)}.\label{eq:3C}
\end{equation}

Panel 2D shows the speed of the Sun relative to the barycenter of
the solar system as used in \citet{Scafetta2010} that can be directly
calculated by the JPL's HORIZONS web-interface;

Panel 2E shows the jerk function (the derivative of the acceleration
vector) of the Sun as defined in \citet{Scafetta2012d}

\begin{equation} \vec{J}_{S}(t)=\dot{\vec{a}}_{S}(t)=\sum_{P=1}^{8}\frac{G~m_{p}~V_{SP}(t)}{R_{SP}^{3}(t)}\left(\frac{\vec{V}_{SP}(t)}{V_{SP}(t)}-3~\frac{\vec{R}_{SP}(t)}{R_{SP}(t)}\right),\label{3E} \end{equation} where
$\vec{V}_{SP}(t)$ is the velocity of the planet P relative to the
Sun. 

The four functions depicted in Panels 2B-2E stress different frequencies
that may be complementary drivers for different TSI periodicities.
For example, as explained above different periodicities can be activated
by alternative physical mechanisms (e.g.: gravitational tides, electromagnetic
interactions, geometrical asymmetry of the solar structure, etc.).

In Figure 2 we observe major frequency peaks at the periods listed
in Tables 1, 2 and 3. For example, the planetary spectral peaks at
the periods of about 0.2 year (cyan bar), 0.25 year (cyan bar), 0.3-0.33
year (yellow bar), 0.4 year (yellow bar), 0.65 year (yellow bar),
0.8 year (yellow bar) and 1.09 year (yellow bar) are very clear, and
are found also in the TSI records (Fig. 2A). The two yellow bars in
each of the spectral panels 2B-2E highlight the two most significant
spectral peaks for each case. 

In particular, by using these yellow bars as guides, we note the following
correspondences with the TSI periodogram depicted in Figure 2A: (1)
comparing Figures 2A and 2B, we find major common spectral peaks at
the spring period of Venus and Jupiter (0.32 year) and other closed
peaks 0.30-0.33 year, and at the spring period of Venus and Earth
(0.80 year); (2) comparing Figures 2A and 2C, we find major common
spectral peaks at about 0.24-0.25 year related to Mercury, at about
0.30-0.33 year related to Venus, Earth and Jupiter, and at 0.5 year
related to half period of the Earth's orbit, which is strong in the
solar z-axis of the tidal component because the inclination of the
Earth's orbit relative to the solar equatorial plane is the largest
among the planets (about $7^{o}$); (3) comparing Figures 2A and 2D,
we find major common spectral peaks at the synodic period of Venus
and Jupiter (0.65 year) and at the synodic period of Jupiter and Earth
(1.09 year); (4) comparing Figures 2A and 2E, we find major common
spectral peaks at the synodic period of Venus and Mercury (0.4 year)
and at the synodic period of Venus and Earth (0.65 year).

The common 0.3-0.33 year harmonic cluster is made of at least three
very close spectral peaks both in the TSI spectrum (Fig. 2A) and in
the planetary spectrum depicted in Figure 2B-2C (see also Tables 1,
2 and 3). Other common peaks are visible, although less important.
For example, the relatively strong spring harmonic between Mercury
and Venus, at about 0.2 year, and the Mercury orbital harmonics at
about 0.24 year are found in Figures 2A-2E. Figure 2A adds a green
bar at 0.5 year that we interpret as produced by the Earth crossing
the solar equator twice a year, which is also clearly visible in Figure
2C. 

The black small bars between panel A2 and B2 indicate all theoretical
spring, orbital and synodic periods expected from the eight planets
of the solar system within the analyzed period band from 0 to 1.2
year reported in Tables 1 and 2. These results clearly suggest that
the TSI record presents spectral fingerprints of multiple planetary
harmonics. Thus, planetary forces likely synchronize and modulate
TSI dynamics probably by means of several mechanisms. 

A statistical confidence value may be estimated in the following way.
The spectral interval from $0$ year to $1.2$ year can be divided
into $20$ non-overlapping bars with a width of $0.06$ year each.
These correspond approximately to the width of the colored bars depicted
in the figures and approximately to the $\pm0.03$ year maximum error
bar among the theoretical frequencies derived from ephemeris calculations.
Figure 1A shows 11 major spectral peaks at a 95\% confidence level,
and at least 8 common spectral peaks are highlighted in Figure 2B-2E
by the colored bars. The probability to find by random chance 8 major
frequency peaks in 8 specific bars on a set of $20$ available frequency
range slots is about $8!12!/20!\approx0.001$\%. 

Note, however, that Figure 2A shows 11 spectral peaks above the 95\%
red-noise confidence level. The three spectral peaks that were not
highlighted with colored bars are at about $0.7$ year, $0.9$ year
and $0.96$ year. These spectral peaks appear also to be present among
the planetary harmonics as evident in the tables: the quasi 0.7-year
cycle appears to be related to the spring period of Venus beating
with the synodic cycle of Earth and Jupiter (see Table 3); the quasi
0.9-year cycle and the cycles at 0.94-1.00 years appear related to
the harmonics associated with Mars (see Table 2). The probability
to find by random chance $11$ major frequency peaks in $11$ specific
bars on a set of $20$ available frequency range slots is about $9!11!/20!\approx0.001$\%.
However, by ignoring the physical attribution of the three additional
spectral peaks, which is more uncertain, it may be better to evaluate
the probability to find by random chance 8 out of 11 major frequency
peaks in 8 specific bars on a set of $20$ available frequency range
slots, which is $(11!9!/20!)*(11!/8!/3!)\approx0.1$\%.

In conclusion, Figures 1 and 2 suggest a high degree of statistical
coherence between the TSI periodogram and the major theoretical expected
planetary frequencies. 

\begin{figure}
\includegraphics[width=1\textwidth]{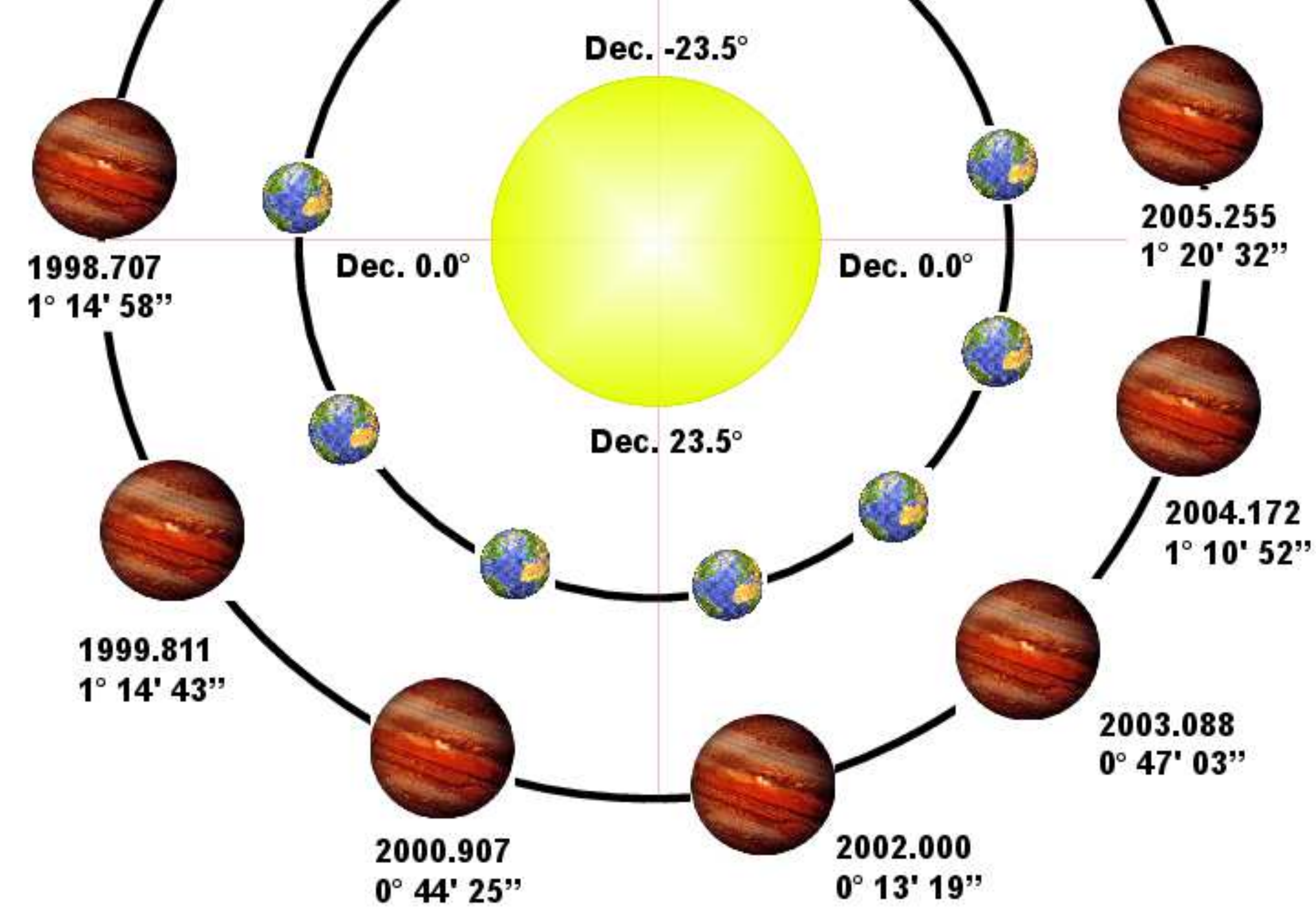}

\caption{The Earth-Jupiter conjunction cycle (1.09-year) during solar cycle
23 maximum (1998-2004). }
\end{figure}

\section{Interpretation of the 1.092 year TSI oscillation as an Earth-Jupiter
conjunction cycle effect}

In this section we propose a dynamical test to determine whether the
TSI oscillations are synchronized to the planetary oscillations and
whether there might be non-linear internal mechanisms causing the
solar response to the external forcing vary in time. Herein, the test
is limited only to the major observed oscillation of the periodogram:
during the maximum of solar cycle 23, $\sim$1.092-year oscillations
are macroscopic and readily visible. If these oscillations are related
to the Earth-Jupiter (1.092 year) synodic cycle, as the spectral analyses
depicted in Figures 1 and 2 suggest, then the TSI record should present
maxima in proximity of the Earth-Jupiter conjunction dates shown in
Figure 3. 

\begin{figure}
\includegraphics[width=1\textwidth,height=0.8\textheight]{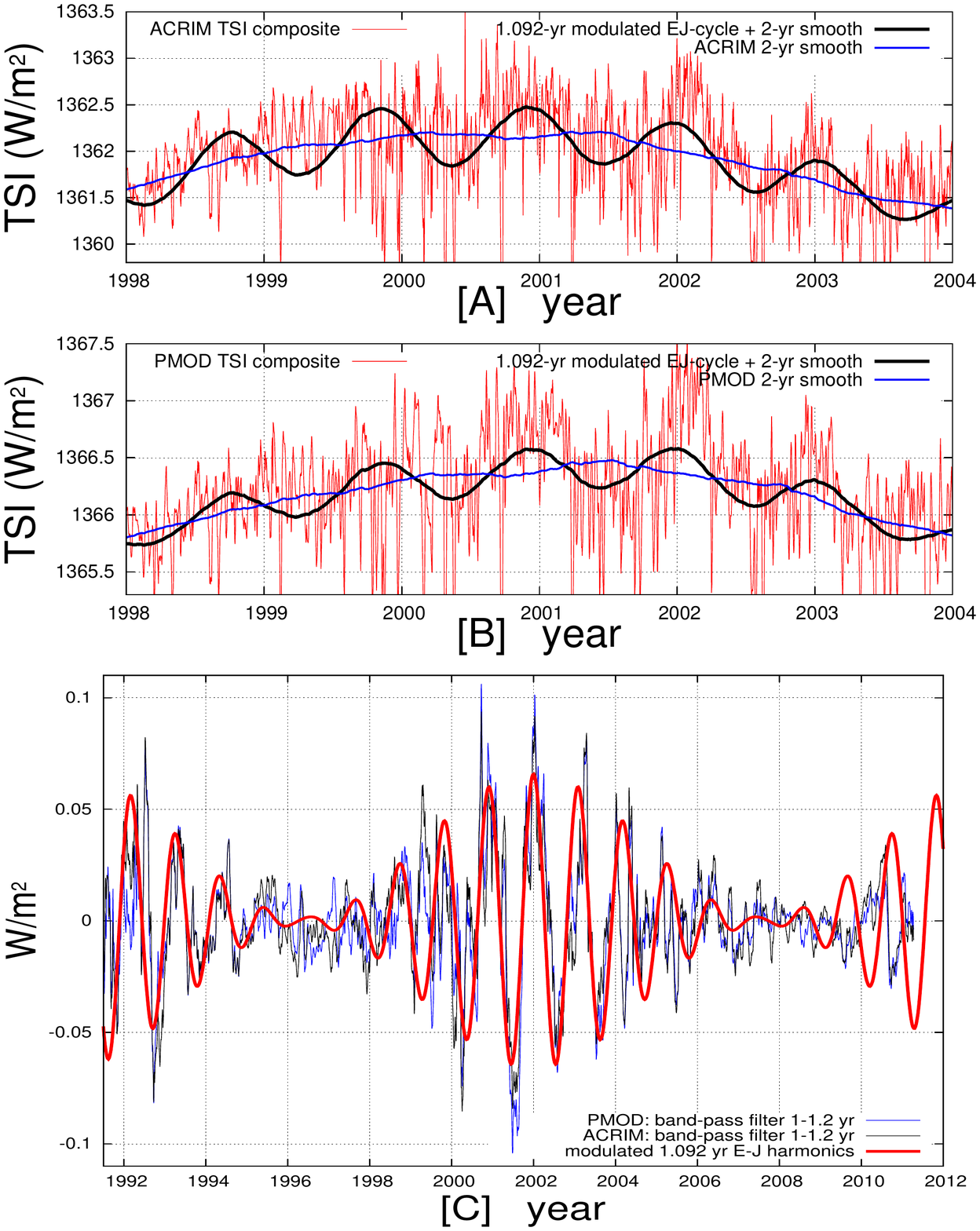}

\caption{{[}A{]} ACRIM and {[}B{]} PMOD TSI composite during solar maximum
23 (1998-2004). The modeled black curve is made as in Figure 5: see
Eqs. 7 and 8. {[}C{]} Moving average algorithms are used to isolate
the 1-1.2 year time scale of PMOD (blue) and ACRIM (black) and compared
against a 1.092-year harmonic function (red) whose amplitude is empirically
modulated on the Schwabe solar cycle. The 1.092-year harmonic function
is chosen in phase with the Earth-Jupiter conjunction cycle. The modeled
curves are approximations used only for visualization purpose. }
\end{figure}

Figures 4A and 4B show ACRIM and PMOD (red curves) against an Earth-Jupiter
(1.092 year) conjunction cycle reconstruction (black curve) for the
1998-2004 period. TSI peaks near the conjunction dates are clearly
seen. The largest peak occurs near 2002 when the conjunction occurs
at a minimum angular separation between Earth and Jupiter (0$^{o}$
13' 19\textquotedbl{}), as expected.

Figure 4C further demonstrates the above issue showing the PMOD (blue)
and ACRIM (black) records band-pass filtered to highlight their 1.0-1.2
year time-scale modulation. The two filtered curves are compared against
a $\sim$1.092-year harmonic function (red):

\begin{equation}
f(t)=g(t)\cos\left[\frac{2\pi(t-2002)}{1.09208}\right],\label{eq:sss}
\end{equation}
whose amplitude $g(t)$ has been modulated on the Schwabe solar cycle.
The maxima of the harmonic function well correspond to the Earth-Jupiter
conjunction times as shown in Figure 3. Note that a Earth-Jupiter
conjunction occurred on Jan/01/2002 and the average synodic period
is 1.09208 year. 

Figure 4C clearly demonstrates that the 1.0-1.2-year time-scale modulation
of the TSI records is well correlated to the 1.092-year Earth-Jupiter
conjunction cycle. The effect is significantly attenuated during solar
minima (1995-1997 and 2007-2009) and increases during solar maxima.
In particular, the figure shows the maximum of solar cycle 23 and
part of the maxima of solar cycles 22 and 24.

Figure 5 compares the ACRIM and PMOD TSI composites since 1978 with
two approximate empirically constructed curves comprised of the Earth-Jupiter
conjunction cycle (1.09-year) as modulated by the 11-year solar cycle.
The modulation assumes that the oscillations are stronger during periods
of solar maxima and are damped during solar minima. We use the following
equations: for ACRIM, 

\begin{equation}
f(t)=S_{A}(t)+0.2(S_{A}(t)-1360.58)\cos(2\pi(t-2002)/1.09208);
\end{equation}
for PMOD,

\begin{equation}
f(t)=S_{P}(t)+0.2(S_{P}(t)-1365.3)\cos(2\pi(t-2002)/1.09208).
\end{equation}
The blue curves are 2-year moving average smooths, $S_{A}(t)$ and
$S_{P}(t)$, for ACRIM and PMOD respectively. A more accurate modeling
of the amplitudes of solar variation using multiple harmonics, their
accurate amplitudes and their couplings is extremely complex, as ocean
tidal studies would suggest \citep{Wang}, and are left to another
study.

Using the modeled black curve as a guide, it is possible to recognize
that the 1.09-year Earth-Jupiter conjunction cycle is present during
the entire TSI composite period since 1978. TSI relative peaks are
also found near other Earth-Jupiter conjunction dates such as in 1979,
1981, 1984, 1990, 1991, 1992, 1993, 1994, 1995, 1998, 2011 and 2012.
The 1979 and 1990 peaks are less evident in PMOD, likely a result
of the PMOD alteration of published Nimbus7/ERB records in 1979 and
1989 \citep{Frohlich,Scafetta2009,Scafetta2011}.

The 1.09-year cycle is expected to be larger at the minimum angular
separation between Earth and Jupiter. Earth's orbital inclination
to the Sun's equator is 7.16$^{o}$ and Jupiter's orbital inclination
is 6.09$^{o}$. Thus, the minimum angular separation occurs when their
conjunctions happen close to the solar equatorial plane. The Earth
crosses the solar equator every year on Dec-08 and on Jun-07. From
1978 to 2020, Jupiter crosses the solar equator on: 1983-Jul-22, 1989-Mar-09,
1995-Jun-02, 2001-Jan-18, 2007-Apr-14, 2012-Nov-27 and 2019-Feb-23.
Indeed, Figure 3 shows larger annual-scale TSI peaks in 1983, 1984,
1995, 2001, and 2002.

Finally, a TSI forecast illustration shown in Figure 5A assumes a
weak solar cycle 24 that peaks in 2014 (suggested by current TSI data)
with a 1.092-year modulation. The 1.09-year periodicity predicts a
TSI decrease from December 2011 to May 2012 followed by a significant
TSI increase from June to December 2012 since the Earth-Jupiter conjunction
cycle angular separation is small (0$^{o}$ 38' 30\textquotedbl{}).
Another strong TSI peak should occur in January 2014 for a similar
reason (angular separation: 0$^{o}$ 09' 34\textquotedbl{}), followed
by another maximum in February 2015. However, many other harmonics
exist and interfere with each other. Therefore, the actual TSI variation
patterns is expected to be more complex than suggested in the depicted
simplified forecast.

\begin{figure}
\begin{centering}
\includegraphics[height=1\textwidth,angle=-90]{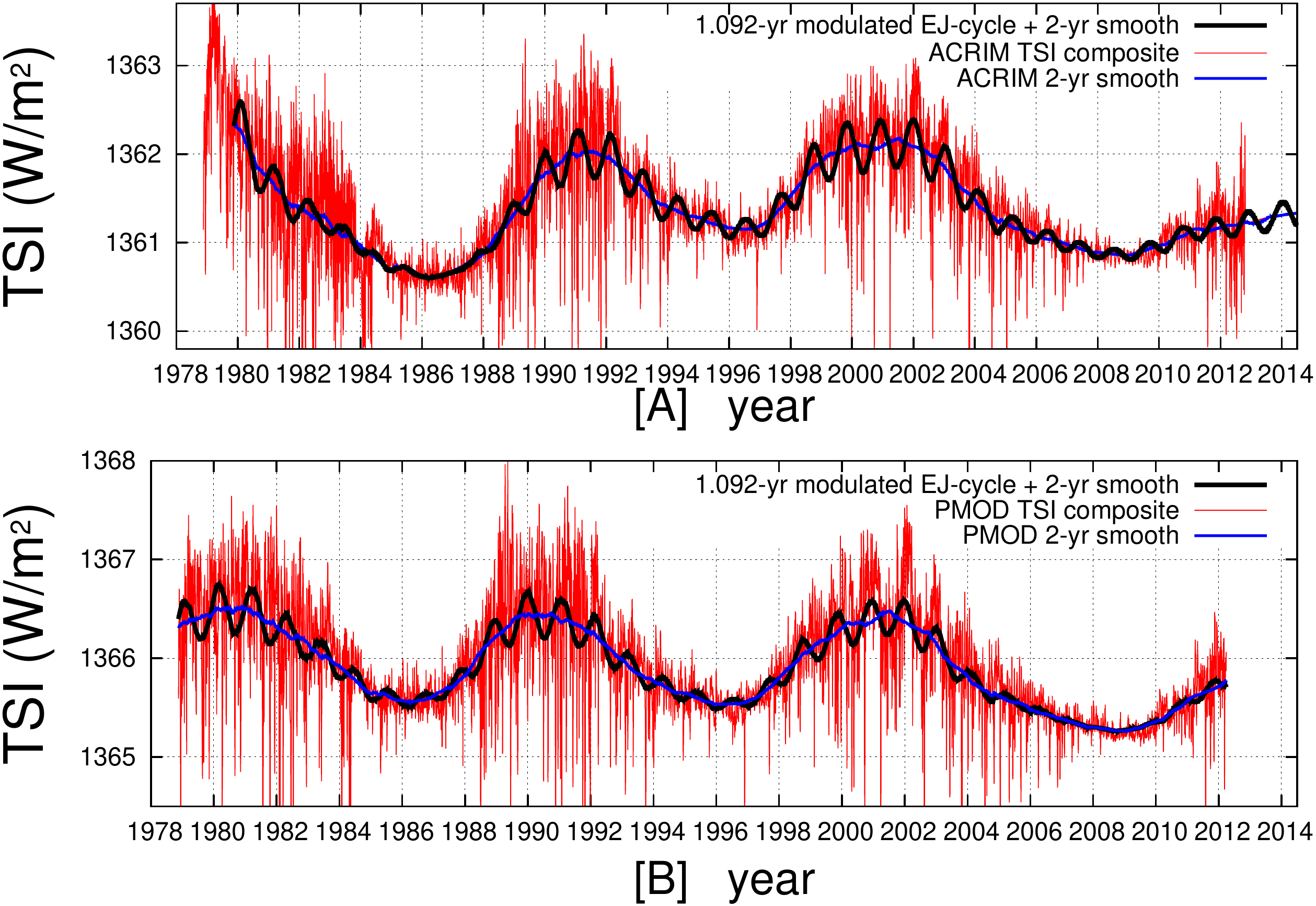}
\par\end{centering}

\caption{{[}A{]} ACRIM and {[}B{]} PMOD TSI satellite composites since 1978
(red). The blue curves are 2-year moving average smooth, $S_{A}(t)$
and $S_{P}(t)$, for ACRIM and PMOD respectively. The black curves
are empirical representations of Earth-Jupiter conjunction 1.092-year
cycle modulated by the 11-year solar cycle. The modeled curves are
approximations used only for visualization purpose: see Eqs. 7 and
8. }
\end{figure}

\section{Conclusions}

We have investigated evidences for planetary-induced solar luminosity
enhancement in the ACRIM and PMOD TSI satellite composites on monthly
to annual timescales. A set of spectral peaks closely matching a set
of planetary gravitational and magnetic theoretical harmonics has
been found. These relatively fast cycles are primarily related to
synodic and spring tidal planetary orbital combinations. The TSI power
spectra present a strong peak around 0.30-0.33 year (109-121 days)
that includes the spring tidal cycle of Jupiter and Venus (the strongest
spring tide within the analyzed frequency band) and other closed planetary
harmonics as indicated in Tables 1 and 2. 

TSI records also show other clustered planetary resonances. In the
45-58 day range there are numerous Mercury spring tides and at 72
days Mercury-Venus. In the 88-91 day range there are Mercury and its
synodic cycles with the Jovian planets. On the 113-120 day range there
are Mercury-Earth synodic cycles and Venus-Jupiter spring tides. At
145 days there is Mercury-Venus synodic cycle. During 225-292 day
periodicities there are numerous Venus related cycles. There is a
major 0.5-year cycle related to the inclination of the ecliptic relative
to the solar equatorial plane. A 1.092-year oscillation related to
the Earth-Jupiter synodic cycle is more readily apparent during solar
cycle maxima. TSI also presents spectral peaks between 0.068 year
(25 days) and 0.111 year (40 days) that are related to the differential
solar rotation periods (see Table 3).

Other researchers have studied the fast oscillations present in alternative
solar indexes and found results compatible with ours. \citet{Rieger}
found the quasi 25-35 days solar rotation cycles and major oscillations
within 138-168 day period in energetic solar flare events, which approximately
corresponds to Mercury\textendash{}Venus synodic cycle. \citet{Pap}
also found evidences for similar oscillations; \citet{Caballero}
reported 58, 78, 89 and 115 day oscillations in cosmic rays time series;
\citet{Kilcik} analyzed the solar flare index and reported frequency
peaks close to 53, 85, 152, 248 days. More recently \citet{Tan} found
evidences of planetary fast harmonics in sunspot and solar microwave
records. 

The 1.09-year Earth-Jupiter conjunction cycle appears to dominate
in the TSI records, in particular during solar cycle maxima. Indeed,
Jupiter's gravitational and magnetic interactions with the Sun are
the strongest among the eight planets of the solar system. The likely
explanation is that the side of the Sun facing Jupiter would be expected
to be the locus of maximum interaction and cause this region to be
slightly brighter causing satellite instruments to detect a slightly
stronger TSI signal when the Earth crosses the Sun-Jupiter conjunction
line. This coalesces with strong hot spots observed on other stars
with closely orbiting giant planets \citep{Shkolnik2003,Shkolnik2005}.

The amplitude dependence of the 1.09-year periodicity on the solar
cycle may indicate that in addition to gravitational tidal interaction
\citep{Wolff,Scafetta2012d}, there may also exist a solar dynamo
amplification mechanism that varies with the Schwabe solar cycle.
There may also exist an enhancement related to the interplanetary
magnetic field structure linking the planets along the Parker spiral
of the stellar wind \citep{Kopp,Gurdemir}.

The harmonics present in the variability of satellite TSI satellite
observations indicate that planetary forcings are likely modulating
solar activity and that both magnetic and gravitational coupling are
involved. Future research will better address the nature of these
couplings, which could provide better forecast capability for solar
activity and climate change. It has been shown in recent studies that
both solar and climate records present strong signatures of planetary
cycles at decadal, secular and millennial time scales \citep{Scafetta2010,Scafetta2012a,Scafetta2012b,Scafetta2012c,Scafetta2012d,Scafetta2013,Scafettaal2013}.

\subsection*{Acknowledgment}

The National Aeronautics and Space Administration supported Dr. Willson
under contracts NNG004HZ42C at Columbia University, Subcontracts 1345042
and 1405003 at the Jet Propulsion Laboratory.

\newpage{}

\begin{table}
\begin{tabular}{ccccccc}
\hline 
Cycle  & Type & P (day)  & P (year)  & min (year) & max (year) & strength\tabularnewline
\hline 
Me & $\nicefrac{1}{2}$ orbital & $44\pm0$ & $0.120\pm0.000$ & $0.120$ & $0.121$ & u\tabularnewline
\hline 
Me \textendash{} Ju & spring & $45\pm9$ & $0.123\pm0.024$ & $0.090$ & $0.156$ & u/w\tabularnewline
\hline 
Me \textendash{} Ea  & spring & $58\pm10$ & $0.159\pm0.027$ & $0.117$ & $0.189$ & u/w\tabularnewline
\hline 
Me \textendash{} Ve & spring & $72\pm8$ & $0.198\pm0.021$ & $0.156$ & $0.219$ & u/w\tabularnewline
\hline 
Me & orbital & $88\pm0$ & $0.241\pm0.000$ & $0.241$ & $0.241$ & u\tabularnewline
\hline 
Me \textendash{} Ju  & synodic & $90\pm1$ & $0.246\pm0.002$ & $0.243$ & $0.250$ & u\tabularnewline
\hline 
Ea & $\nicefrac{1}{4}$ orbital & $91\pm3$ & $0.250\pm0.000$ & $0.250$ & $0.250$ & w\tabularnewline
\hline 
Ve & $\nicefrac{1}{2}$ orbital & $112.5\pm0$ & $0.307\pm0.000$ & $0.307$ & $0.308$ & w\tabularnewline
\hline 
Me \textendash{} Ea  & synodic & $116\pm9$ & $0.317\pm0.024$ & $0.290$ & $0.354$ & u/w\tabularnewline
\hline 
Ve \textendash{} Ju & spring & $118\pm1$ & $0.324\pm0.003$ & $0.319$ & $0.328$ & u\tabularnewline
\hline 
Ea & $\nicefrac{1}{3}$ orbital & $121\pm7$ & $0.333\pm0.000$ & $0.333$ & $0.333$ & w\tabularnewline
\hline 
Me \textendash{} Ve  & synodic & $145\pm12$ & $0.396\pm0.033$ & $0.342$ & $0.433$ & u/w\tabularnewline
\hline 
Ea & $\nicefrac{1}{2}$ orbital & $182\pm0$ & $0.500\pm0.000$ & $0.500$ & $0.500$ & u\tabularnewline
\hline 
Ea \textendash{} Ju & spring & $199\pm3$ & $0.546\pm0.010$ & $0.531$ & $0.562$ & w\tabularnewline
\hline 
Ve & orbital & $225\pm0$ & $0.615\pm0.000$ & $0.615$ & $0.615$ & w\tabularnewline
\hline 
Ve \textendash{} Ju  & synodic & $237\pm1$ & $0.649\pm0.004$ & $0.642$ & $0.654$ & u\tabularnewline
\hline 
Ve \textendash{} Ea  & spring & $292\pm3$ & $0.799\pm0.008$ & $0.786$ & $0.810$ & u/w\tabularnewline
\hline 
Ea & orbital & $365.25\pm0$ & $1.000\pm0.000$ & $1.000$ & $1.000$ & w\tabularnewline
\hline 
Ea \textendash{} Ju  & synodic & $399\pm3$ & $1.092\pm0.009$ & $1.082$ & $1.104$ & u/s\tabularnewline
\hline 
Ea \textendash{} Ve & synodic & $584\pm6$ & $1.599\pm0.016$ & $1.572$ & $1.620$ & u\tabularnewline
\hline 
\end{tabular}\caption{List of the major theoretical expected harmonics associated with planetary
orbits within 1.2 year period. $P$ is the period of the harmonic:
see also Eq. \ref{eq:1}. Mercury (Me), Venus (Ve), Earth (Ea), Jupiter
(Ju). The variability ranges are based on ephemeris calculations.
The strength (last column) is qualitatively estimated as: uncertain/strong
(u/s); weak (w); uncertain (u); uncertain/weak (u/w). }
\end{table}
\begin{table}
\begin{tabular}{ccccc}
\hline 
Cycle  & Type & P (year)  & Type & P (year) \tabularnewline
\hline 
Me \textendash{} Ne & spring & $0.1206$ & synodic & $0.2413$\tabularnewline
\hline 
Me \textendash{} Ur  & spring & $0.1208$ & synodic & $0.2416$\tabularnewline
\hline 
Me \textendash{} Sa & spring & $0.1215$ & synodic & $0.2429$\tabularnewline
\hline 
Me \textendash{} Ma & spring & $0.1382$ & synodic & $0.2763$\tabularnewline
\hline 
Ve \textendash{} Ne & spring & $0.3088$ & synodic & $0.6175$\tabularnewline
\hline 
Ve \textendash{} Ur & spring & $0.3099$ & synodic & $0.6197$\tabularnewline
\hline 
Ve \textendash{} Sa & spring & $0.3142$ & synodic & $0.6283$\tabularnewline
\hline 
Ve \textendash{} Ma & spring & $0.4571$ & synodic & $0.9142$\tabularnewline
\hline 
Ea \textendash{} Ne & spring & $0.5031$ & synodic & $1.006$\tabularnewline
\hline 
Ea \textendash{} Ur & spring & $0.5060$ & synodic & $1.0121$\tabularnewline
\hline 
Ea \textendash{} Sa & spring & $0.5176$ & synodic & $1.0352$\tabularnewline
\hline 
Ea \textendash{} Ma & spring & $1.0676$ & synodic & $2.1352$\tabularnewline
\hline 
Ma & $\nicefrac{1}{2}$ orbital & $0.9405$ & orbital & $1.8809$\tabularnewline
\hline 
Ma \textendash{} Ne & spring & $0.9514$ & synodic & $1.9028$\tabularnewline
\hline 
Ma \textendash{} Ur & spring & $0.9621$ & synodic & $1.9241$\tabularnewline
\hline 
Ma \textendash{} Sa & spring & $1.0047$ & synodic & $2.0094$\tabularnewline
\hline 
Ma \textendash{} Ju & spring & $1.1178$ & synodic & $2.2355$\tabularnewline
\hline 
Ju & $\nicefrac{1}{2}$ orbital & $5.9289$ & orbital & $11.858$\tabularnewline
\hline 
Ju \textendash{} Ne & spring & $6.3917$ & synodic & $12.783$\tabularnewline
\hline 
Ju \textendash{} Ur & spring & $6.9067$ & synodic & $13.813$\tabularnewline
\hline 
Ju \textendash{} Sa & spring & $9.9310$ & synodic & $19.862$\tabularnewline
\hline 
Sa & $\nicefrac{1}{2}$ orbital & $14.712$ & orbital & $29.424$\tabularnewline
\hline 
Sa \textendash{} Ne & spring & $17.935$ & synodic & $35.870$\tabularnewline
\hline 
Sa \textendash{} Ur & spring & $22.680$ & synodic & $45.360$\tabularnewline
\hline 
Ur & $\nicefrac{1}{2}$ orbital & $41.874$ & orbital & $83.748$\tabularnewline
\hline 
Ur \textendash{} Ne & spring & $85.723$ & synodic & $171.45$\tabularnewline
\hline 
Ne & $\nicefrac{1}{2}$ orbital & $81.862$ & orbital & $163.72$\tabularnewline
\hline 
\end{tabular}\caption{List of additional average theoretical expected harmonics associated
with planetary orbits: Mercury (Me), Venus (Ve), Earth (Ea), Mars
(Ma), Jupiter (Ju), Saturn (Sa), Uranus (Ur), Neptune (Ne). See Eq.
\ref{eq:1}.}
\end{table}

\begin{table}
\begin{tabular}{ccccc}
Cycle  & Type & P (year)  & Type & P (year) \tabularnewline
\hline 
Sun  & equ-rot & 0.068  & pol-rot & 0.094\tabularnewline
\hline 
Sun \textendash{} Ju & equ-rot & 0.068 & pol-rot & 0.095\tabularnewline
\hline 
Sun \textendash{} Ea & equ-rot & 0.073 & pol-rot & 0.104\tabularnewline
\hline 
Sun \textendash{} Ve & equ-rot & 0.076 & pol-rot & 0.111\tabularnewline
\hline 
Sun \textendash{} Me & equ-rot & 0.095 & pol-rot & 0.154\tabularnewline
\hline 
Me \textendash{} (Ju \textendash{} Sa) & spring & 0.122  & synodic & 0.244 \tabularnewline
\hline 
Me \textendash{} (Ea \textendash{} Ju) & spring & 0.155  & synodic & 0.309 \tabularnewline
\hline 
Ve \textendash{} (Ju \textendash{} Sa) & spring & 0.317  & synodic & 0.635\tabularnewline
\hline 
Ea \textendash{} (Ju \textendash{} Sa) & spring & 0.527  & synodic & 1.053\tabularnewline
\hline 
Ve \textendash{} (Ea \textendash{} Ju) & spring & 0.704  & synodic & 1.408 \tabularnewline
\hline 
\end{tabular}\caption{Partial list of additional theoretical harmonics. First rows: solar
equatorial (equ-) and polar (pol-) rotation cycles relative to the
fixed stars and to the four major tidal planets calculated using Eq.
\ref{eq:1}. Other rows report the spring and synodic periods of Mercury,
Venus and Earth relative to the synodic periods of Jupiter and Saturn,
and Earth and Jupiter. The latter periods are calculated using Eq.
\ref{eq:1-1}.}
\end{table}

\end{document}